\documentclass[10pt,final,journal,oneside,twocolumn]{IEEEtran}

\usepackage{cite}
\usepackage{graphicx}
\usepackage{subfigure}
\usepackage{epsfig}
\usepackage[update,prepend]{epstopdf}
\usepackage{amssymb,amsmath}
\usepackage{times}

\newtheorem{theorem}{\textbf{Theorem}}

\newtheorem{proposition}[theorem]{\textbf{Proposition}}
\newtheorem{corollary}[theorem]{\textbf{Corollary}}

\newcounter{MYtempeqncnt}

\begin{document}

\title{On the Broadcast Latency in Finite Cooperative Wireless Networks}

\author{Anvar~Tukmanov,~\IEEEmembership{Student Member,~IEEE,}
				Zhiguo~Ding,~\IEEEmembership{Member,~IEEE,}
				Said~Boussakta,~\IEEEmembership{Senior Member,~IEEE,}
				and Abbas~Jamalipour,~\IEEEmembership{Fellow,~IEEE}		
\thanks{Manuscript received August 16, 2011; revised November 2, 2011; accepted January 5, 2012. This work was supported in part by the Royal Society of Engineering Distinguished Visiting Fellowship Scheme Grant 10508/366 and the UK EPSRC Grant EP/I037423/1. The associate editor coordinating the review of this paper and approving it for publication was Prof. Steven D. Blostein.}
\thanks{A. Tukmanov, Z. Ding and S. Boussakta  are with the School of Electrical, Electronic, and Computer Engineering, Newcastle University, NE1 7RU, UK (e-mail: \{anvar.tukmanov, zhiguo.ding, said.boussakta\}@ncl.ac.uk).}
\thanks{A. Jamalipour is with the School of Electrical and Information Engineering, University of Sydney, NSW, 2006, Australia (e-mail: a.jamalipour@ieee.org).}
}

%% The paper headers
%\markboth{Journal of \LaTeX\ Class Files,~Vol.~6, No.~1, January~2007}%
%{Shell \MakeLowercase{\textit{et al.}}: Bare Demo of IEEEtran.cls for Journals}

\maketitle

\begin{abstract}
%\boldmath
The aim of this paper is to study the effect of cooperation on system delay, quantified as the number of retransmissions required to deliver a broadcast message to all intended receivers. Unlike existing works on broadcast scenarios, where distance between nodes is not explicitly considered, we examine the joint effect of small scale fading and propagation path loss. Also, we study cooperation in application to finite networks, i.e. when the number of cooperating nodes is small. Stochastic geometry and order statistics are used to develop analytical models that tightly match the simulation results for non-cooperative scenario and provide a lower bound for delay in a cooperative setting. We demonstrate that even for a simple flooding scenario, cooperative broadcast achieves significantly lower system delay. 
\end{abstract}


\IEEEpeerreviewmaketitle

\section{Introduction}
Cooperation between wireless nodes has gained wide attention as it allows trading extra spatial degree of freedom for reduced outage probability, increased capacity or lower power consumption \cite{erk03, lan04}. Several works have studied the effects of node cooperation on broadcast scenarios, where a source wants to deliver a message to multiple destinations \cite{wornell06, sirkeci07}. However, until recently \cite{leung10}, explicit network geometry and associated path loss (PL) effects were either ignored, or considered for some special network settings. Namely, in \cite{wornell06} a framework facilitating outage probability analysis is proposed for cooperative multicast schemes in presence of Rayleigh fading in \emph{dense wireless networks}, i.e. when the number of nodes is large. In \cite{sirkeci07} PL between nodes has been considered in context of power efficiency of broadcast strategies again in the context of dense networks.

The aim of this work is to analyze the transmission delay for broadcast in networks with \emph{finite} number of highly mobile nodes and to compare performance of cooperative broadcast to conventional non-cooperative  protocol. We also aim a realistic description of channel conditions, and use stochastic geometry to account for path loss in addition to small scale Rayleigh fading. Stochastic geometry tools that allow joint treatment of both channel and location randomness have been studied in-depth for infinite networks in \cite{ha08}. However, their potential has not been applied widely to cooperation, and only initial analysis of inter-node distances for finite networks was presented in \cite{finite}.

Our main contribution is in the methodology that allows analysis of \emph{finite} random networks, incorporating both channel and node location randomness. We show that tractable analytical results, that match closely to Monte-Carlo simulations, can be derived for realistic fading and path loss conditions. Our results demonstrate that cooperative schemes can achieve significantly lower system delay compared to non-cooperative broadcast. We also provide simulation results to emphasize the impact of network size and node density on performance of broadcast schemes. Namely, performance of the non-cooperative scheme depends primarily on network size, whereas the cooperative scheme is more sensitive to node density.

\section{System model}\label{sec:model}
In this section, we briefly describe broadcast protocols which have been studied, discuss required stochastic geometry theory and transmission latency metric. 
A circular cell structure is considered with the source node located in the center. The source aims to deliver a common message to all $N$ randomly located and highly mobile wireless nodes.
\subsection{Broadcast protocols}
We will adopt broadcast schemes previously used in \cite{wornell06} as a basis for studied protocols. In \textit{cooperative broadcast} the source broadcasts a message in the first time slot, and continues until at least one receiver receives it correctly. After that the source remains silent, while all successful receivers cooperate by simultaneously retransmitting the message to remaining nodes in subsequent time slots until all nodes receive the message. Any receiver is assumed to process signals only from the nearest transmitter. For \textit{non-cooperative broadcast} -- only source transmits and all remaining nodes listen.

Note that non-orthogonal transmissions used in this work may lead to performance degradation due to multipath arrivals of the same message. This effect is accounted for by Rayleigh fading. Similar setting has been used in \cite{sirk06}, where cooperatively transmitted signals were additionally coherently combined at destination. Although coherent combining can improve performance, it is difficult to implement due to system overhead. Therefore, only the nearest transmitter's signal is considered in this paper as its power is expected to dominate the rest.

A high mobility (HM) model \cite{ha10mob} is used in this paper, where node positions change randomly and independently in each time slot. In practice, node locations may be dependent and follow certain law. Such mobility models could be described by more advanced PP constructions, for example, the hard-core PP \cite{stoyan08} could be used to capture the effect of minimal expected node speed. A setting with static nodes was studied in \cite{zhe08}, where Voronoi tessellations were used to derive bounds on broadcast capacity. System performance will definitely depend on the chosen mobility model, however the impact that different mobility models have is beyond the scope of this paper.
\subsection{Point processes for node location description}
\begin{figure}[!t]
\begin{center}
\includegraphics[width=\columnwidth]{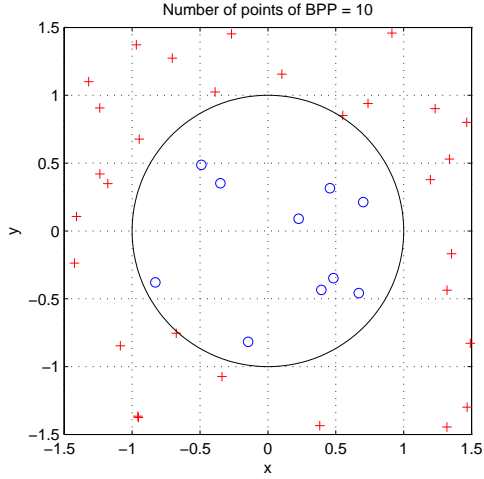}
\caption{A realisation of Binomial point process with $N=10$ nodes in a circular cell.}
\label{fig:BPP_plot}
\end{center}
\end{figure}
Point processes provide a mechanism to analyze interactions between spatially distributed objects, such as wireless nodes \cite{stoyan08}, which allows general yet simple results compared to conventional methods \cite{SGeom}.

Of particular importance is the Poisson point process (PPP), characterised by two fundamental properties \cite{stoyan08}: (1) the probability of having exactly $N$ nodes in a Borel subset set $A$ of some observation window $W$ is Poisson distributed with parameter $\lambda \cdot  \nu_d(A)$; (2) the counts of points in disjoint subsets of $W$ are independent. Here $\lambda$ is the node density and $\nu_d(A)$ is Lebesgue measure (i.e. length, area or volume of $A$) and $W$ is a geometrical construction in space, where the point process is observed. For our purposes\footnote{Interested reader can find a general treatment on types and interactions of sets in  \cite{stoyan08}.}, $W=b_2(o,R)$ denotes a 2-dimensional disk $b_2$ of radius $R$ centered at origin $o$, and $A$ is a geometrical shape, entirely contained in $W$.

While PPPs allow getting valuable results for asymptotic scenarios, once the number of nodes in the network is known, the PPP model becomes inappropriate since the node counts in different locations become dependent (see \cite[p.27]{stoyan08} and \cite{finite}). Therefore we further work with a binomial point process (BPP), characterised by having exactly $N$ nodes located randomly and independently in $W$ (Fig.~\ref{fig:BPP_plot}). Since HM is assumed, each time slot faces a new realization of the BPP, and we can treat the nodes that have received the message correctly and those that have not as two separate BPPs.

\subsection{Latency metric and signal model}
The metric of our interest is the number of transmission attempts, or time slots, $K$ required to deliver a broadcast message to all $N$ nodes within the observation window $W$. The expected number of required transmissions $\bar{K}$ can be found as
\begin{equation}\label{eq:numb_tx}
	\bar{K}=\sum\limits_{k=1}^{\infty} k \cdot P(K=k|N),
\end{equation}
where $P(K=k|N)$ denotes the probability that, conditioned on having $N$ nodes in total, exactly $k$ transmissions will be sufficient to reach all nodes.

We consider joint effect of small scale Rayleigh fading and propagation loss due to respective node locations. In particular, the signal received by node $i$ from transmitter $j$ can be expressed as 
\begin{equation}
	y_i=h_{ij}\sqrt{P_{tx} \cdot l(r_{ij})}\cdot x_j +n_w,
\end{equation}
where $x_j$ is the message, transmitted by  $j$-th node, $P_{tx}$ -- transmission power; $n_w$ is zero-mean AWGN with variance $\sigma^2_w$; $h_{ij} \sim\mathcal{N}(0;1)$  represents the effect of small scale fading; $l(r_{ij})=\left(1+r^{\alpha}_{ij}\right)^{-1}$ is the path loss function \cite{sirkeci07}, $r_{ij}$ is the distance between nodes $i$ and $j$ and $\alpha$ is the path loss exponent.

A message is considered to be successfully delivered if the capacity of communication channel between the transmitting and receiving node is sufficient for transmission at required data rate $\mathcal{R}$:
\begin{equation}
	\log\left(1+ \frac{|h_{ij}|^2 P_{tx}}{\left(1+r^{\alpha}_{ij}\right)\sigma_w^2} \right) \geq \mathcal{R}.
\end{equation}
Probability of such event can be described as probability of success and expressed as
\begin{align}
	P_s=\text{Pr}\left(\frac{|h_{ij}|^2}{1+r^{\alpha}_{ij}} \geq \theta\right),
\end{align}
where $h_{ij}$ and $r_{ij}$ are random variables and $\theta=\frac{\sigma_w^2\left(2^\mathcal{R}-1\right)}{P_{tx}}$ is the threshold for successful reception.

\section{Latency analysis for non-cooperative broadcast}
In order to find the expected number of required transmissions $\bar{K}$ we need to estimate the probability $P(K=k|N)$, described earlier. Let $S_k$, $k \in [1,K]$, be the random variable, representing the number of successful nodes obtained as a result of $k$-th transmission stage, and $T_k=\sum_{i=1}^{k} S_i$ be the total number of successful nodes after $k$ transmissions. Then the broadcast process completes when $T_K=\sum_{i=1}^{K} S_i=N$. Therefore, the metric of our interest can be expressed as sum of probabilities of all possible outcomes of the $k$ transmission stages which lead to the completion of the process in exactly $K=k$ time slots, as can be seen from \eqref{eq:k_tx_ref} on top of this page.
\begin{figure*}[!t]
\normalsize 
\setcounter{MYtempeqncnt}{\value{equation}}
\setcounter{equation}{4}
\begin{equation}\label{eq:k_tx_ref}
\begin{aligned}
P_{k}^{nc}&= P(K=k|N) = P(T_k=N |N) = \sum\limits_{(T_{k-1},S_k)\in c_k} \left\{P_1(S_k|N-T_{k-1}) \cdot P(T_{k-1}| N)\right\} =\sum\limits_{(T_{k-1},S_k)\in c_k } \Biggr\{ P_1(S_k|N-T_{k-1}) \\
& \times \sum\limits_{(T_{k-2},S_{k-1})\in c_{k-1}} \biggr\{P_1(S_{k-1}|N-T_{k-2}) \cdot \dotso \cdot \sum\limits_{(T_{1},S_{2})\in c_{2}} \left\{P_{1}(S_{2}|N-T_{1}) \cdot P_1(T_1|N) \right\} \dotso  \biggr\} \Biggr\},
\end{aligned}
\end{equation}
\setcounter{equation}{\value{MYtempeqncnt}} 
\hrulefill 
\vspace*{4pt}
\end{figure*}
The conditions of summations $c_i, i \in [2,k]$ in \eqref{eq:k_tx_ref} are given as:
\setcounter{equation}{5} 
\begin{equation}\label{eq:sum}
\begin{aligned}
c_k=\bigr\{ &T_{k-1} \in [0,N-1],S_k \in [1,N]: T_{k-1}+S_k=N\bigr\}; \\
c_{k-1}=\bigr\{ &T_{k-2} \in [0,T_{k-1}],S_{k-1}\in [0,T_{k-1}]:\\ &T_{k-2}+S_{k-1}=T_{k-1}\bigr\};
\end{aligned}
\end{equation}
Thus to evaluate \eqref{eq:k_tx_ref}, we need to calculate $P_1(S_i|N-T_{i-1})$, which denotes the probability of getting exactly $S_i$ successful nodes out of $N-T_{i-1}$ as a result of source transmission. To achieve this we will first obtain the distribution function of the compound random variable $Z=\frac{|h_{ij}|^2}{1+r^{\alpha}_{ij}}$ for the independent and identically distributed (i.i.d.) receivers and then use order statistics.

\begin{proposition}[Joint fading-path loss distribution]\label{JointDistr} 
The cumulative distribution function (CDF) of the compound random variable $Z=|h|^2 l(r)$, where $|h|^2$ is the amplitude of Rayleigh fading coefficient and $l(r)=\left(1+r^{\alpha}\right)^{-1}$ is the path loss function, is given by
\begin{align}\label{eq:fz}
	F_Z\left(\theta\right)=1-\frac{\delta e^{-\theta}}{R^d \theta^{\delta}}\gamma(\delta, R\theta),
\end{align}
with $\gamma(\cdot)$ denoting the lower incomplete Gamma function, $d$ is the number of dimensions and $\delta=d/\alpha$ is used for brevity.
\end{proposition}
\begin{IEEEproof}
We can express the CDF of the random variable $Z=\frac{|h|^2}{1+r^{\alpha}}$ as $F_Z (\theta)=\text{Pr}(|h|^2 < \theta \left(1+r^{\alpha}\right))$.
Let us consider components of Z individually and distinguish three cases: $\theta \in (0,\infty)$, $\theta=0$ and $\theta=\infty$.\\
The amplitude of Rayleigh fading coefficient is a Chi-square distributed random variable with two degrees of freedom, which is equivalent to the exponential distribution, i.e. $|h|^2 \sim Exp(1)$. 
The points of the BPP inside $W$ are i.i.d. distributed with common density function (see \cite{ha08} and \cite{vere}) 
\begin{align}\label{eq:f_iid}
	f_{r^{\alpha}}(y)=\frac{\lambda(y)}{\Lambda(W)}=\frac{\delta y^{\delta-1}}{ R^{d}}, y \in \left[0,R^{\alpha}\right],
\end{align}
where  $\Lambda(W)=\#\left\{\Phi \cap W\right\}=\int\limits_{B} \lambda(w) \text{d}w$ is the counting measure for the originating PPP, with $\#\{\cdot\}$ denoting the number of elements in a set; $\lambda(w)$ is the intensity of points in $d$ dimensions, and $\delta=\frac{d}{\alpha}$ is used for compactness.
 We can now find $F_Z\left(\theta\right)$ for the three regions of $\theta$. For $\theta \in (0,\infty)$:
\begin{equation}\label{eq:CDF}
\begin{aligned}
&F_Z (\theta)=
\text{Pr}(|h|^2 < \theta \left(1+r^{\alpha}\right))\\
&=\int\limits_{y=0}^{y=R^{\alpha}} f_{r^{\alpha}}(y) \int\limits_{x=0}^{x=\theta (1+y)} f_{|h|^2}(x) \text{d}x \text{d}y \\
&=\int\limits_{0}^{R^{\alpha}} \frac{\delta y^{\delta-1}}{R^d} \left(1-e^{-\theta(1+y)}\right)\text{d}y\\
&=\underbrace{\frac{\delta}{R^d}\int\limits_{0}^{R^{\alpha}} y^{\delta-1} \text{d}y}_{=1} - \frac{\delta e^{-\theta}}{R^d}\int\limits_{0}^{R^{\alpha}} y^{\delta - 1} \cdot e^{-\theta y} \text{d}y\\
&=1-\frac{\delta e^{-\theta} }{R^d \theta^{\delta}} \gamma(\delta, R^{\alpha} \theta).
\end{aligned}
\end{equation}
The cases of $\theta=0$ and $\theta=\infty$ correspond to the events of the path gain being less than zero or less than infinity. Therefore we can write:
\begin{equation}
F_Z (0)= \text{Pr}(|h|^2 < 0)=0, \quad F_Z (\infty)= \text{Pr}(|h|^2 < \infty)=1. \nonumber
\end{equation}
\end{IEEEproof}

\begin{figure*}[!th]
\normalsize 
\setcounter{MYtempeqncnt}{\value{equation}}
\setcounter{equation}{13}
\begin{equation}\label{eq:k_tx}
\begin{aligned}
	P_{k}^{c}&= P(T_k=N | N)=\sum\limits_{(T_{k-1},S_k)\in c_k} \left\{P_k(S_k|N-T_{k-1}) \cdot P(T_{k-1}| N)\right\}= \sum\limits_{(T_{k-1},S_k)\in c_k} \Biggr\{P_k(S_k|N-T_{k-1}) \\
	&\times \sum\limits_{(T_{k-2},S_{k-1})\in c_{k-1}} \biggr\{P_{k-1}(S_{k-1}|N-T_{k-2}) \cdot \dotso \cdot \sum\limits_{(T_{1},S_{2})\in c_{2}} \left\{P_{2}(S_{2}|N-T_{1}) \cdot P_1(T_1|N) \right\} \dotso \biggr\} \Biggr\},
\end{aligned}
\end{equation}
\setcounter{equation}{\value{MYtempeqncnt}} 
\hrulefill 
\vspace*{4pt}
\end{figure*}

Next we use the above result to derive the probability $P_1(S_i|N-T_{i-1})$ that there are exactly $S_i$ successful nodes out of $N-T_{i-1}$ receivers. For simplicity, we will denote $N_r=N-T_{i-1}$
\begin{corollary}[Order statistics]
The conditional probability of having exactly $S_i$ successful nodes out of $N_r$ receivers is
\begin{equation}\label{eq:p1_noncoop}
\begin{aligned}
	\text{Pr}(S_i|N_r)&={N_r \choose S_i} \left(1-\frac{\delta e^{-\theta} }{R^d \theta^{\delta}}\gamma(\delta, R^{\alpha}\theta)\right)^{N_r-S_i}\\ &\times \left(\frac{\delta e^{-\theta} }{R^d \theta^{\delta}}\gamma(\delta, R^{\alpha}\theta)\right)^{S_i}
\end{aligned}
\end{equation}
\end{corollary}
\begin{IEEEproof}
Given $N_r$ nodes in the process $\Psi$, the probability that there are exactly $S_i$ successful receivers after the first transmission attempt can be expressed as:
\begin{equation}\label{eq:Prob_stage1}
\begin{aligned}
	\text{Pr}(S_i|N_r)=\text{Pr}(Z_{(N_r-S_i)}<\theta, Z_{(N_r-S_i+1)} \geq \theta),
\end{aligned}
\end{equation}
where the terms $Z_{(1)}<Z_{(2)}<\dotso < Z_{(N_r-S_i)} < Z_{(N_r-S_i+1)} < \dotso <Z_{(N_r)}$ correspond to ordered realizations of the random variable $Z$, for which probability distributions are known.
Using order statistics and Proposition \ref{JointDistr}  we can rewrite \eqref{eq:Prob_stage1} as
\setcounter{equation}{9}
\begin{equation}\label{eq:p1}
\begin{aligned}
	&\text{Pr}(S_i|N_r)=\int\limits_{0}^{\theta} \int\limits_{\theta}^{\infty} f_{Z_{(N_r-S_i)},Z_{(N_r-S_i+1)}}(u,v) dv du\\
	&={N_r \choose S_i+1} \int\limits_{0}^{\theta} \left(F_Z(u)\right)^{N_r-S_i-1} \text{d}F_Z(u) \\ & \times\int\limits_{\theta}^{\infty} \left(1-F_Z(v)\right)^{S_i-1} \text{d}F_Z(v)\\
		&={N_r \choose S_i} \left(F_Z (\theta)\right)^{N_r-S_i} \left(1-F_Z (\theta)\right)^{S_i},
\end{aligned}
\end{equation}
\end{IEEEproof}
where $F_Z (\theta)$ is defined in \eqref{eq:fz}.
If $\delta=1$, we get $\gamma(\delta, R^{\alpha}\theta)=\gamma(1, R^{\alpha}\theta)=1-e^{-R^d\theta}$ and:
\begin{equation}\label{eq:p1_noncoop_sp}
\begin{aligned}
	\text{Pr}(S_i|N_r)&={N_r \choose S_i} \frac{1}{(R^d\theta)^{N_r}} \left(e^{-\theta} - e^{-\theta(R^d+1)}\right)^{S_i}\\
	&\times \left(R^d\theta-e^{-\theta} + e^{-\theta(R^d+1)}\right)^{N_r-S_i}.
\end{aligned}
\end{equation}
Substitution of \eqref{eq:p1_noncoop} or \eqref{eq:p1_noncoop_sp} into \eqref{eq:k_tx_ref} gives the desired expected number of required transmissions for non-cooperative broadcast.

\section{Latency analysis for cooperative broadcast}
\subsection{General setting}
Following the same line of reasoning as for \eqref{eq:k_tx_ref}, we find all possible combinations of outcomes of the $k$ transmission stages leading to $T_k=N$ using \eqref{eq:k_tx} on top of this page, where the summation conditions $c_i$ are identical to \eqref{eq:sum}. Here the probability $P_i(S_i|N-T_{i-1})$ denotes the chance to get exactly $S_i$ successful nodes as a result of $i$-th transmission given $N-T_{i-1}$ remaining receivers. The case of $i=1$ corresponds to non-cooperative transmission by the source, which has been analyzed in previous section.\footnote{It has to be mentioned that there is a non-zero probability that a number of source transmissions will not reach any of the receivers, meaning that cooperative stage cannot start. Equation \eqref{eq:k_tx} accounts for such events, with the probability of throttle transmissions $P_i(S_i|N-0)$ equivalent to the probability of getting zero successful nodes in a non-cooperative scenario $P_1(S_1=0|N)$.} We will further estimate $P_i=P_i(S_i|N-T_{i-1})$ for cooperative stages.

\subsection{Estimation of $P_i=P_i(S_i|N-T_{i-1})$}
At the $i$-th stage the message is retransmitted by all successful receivers, originated in $(i-1)$ previous transmission stages. We are interested in the event when exactly $S_i$ of the receivers successfully receive the message while $(N-T_{i-1}-S_i)$ do not.

Any particular receiver can be treated as a reference point of a BPP of transmitters, containing $T_{i-1}$ nodes. As  each receiver is restricted to processing signals only from the nearest transmitter, we would like to find corresponding \textit{distribution of SNR under joint effect of fading and path loss}.
Associated difficulty is that the BPP process of the transmitters becomes anisotropic once the observation point is shifted from the origin of circular cell. However, we will concentrate on the isotropic scenario with the reference point located at the origin, which will give an approximation of performance, keeping derivations feasible. 

Since the distributions of distances from all receivers to transmitters are assumed to be identical, we can express the probability $P_i(S_i |N- T_{i-1})$ of getting exactly $S_i$ successful nodes given $N-T_{i-1}$ remaining receivers as
\setcounter{equation}{14}
\begin{align}\label{eq:pk}
	P_i(S_i|N-T_{i-1})= {N-T_{i-1} \choose S_i} P_s^{S_i} \left(1-P_s\right)^{N-T_{i-1}-S_i},
\end{align}
where  $P_s=P\left(\frac{|h|^2}{1+r^{\alpha}} \geq \theta\right)$ is the probability of successful communication for transmitter-receiver pair.
Next we will estimate $P_s$.
\begin{proposition}[Probability of success]
Under an isotropic BPP assumption, the probability of successful communication between a receiver and its nearest transmitter under the joint effect of Rayleigh fading and path loss of $l(r)=\left(1+r^{\alpha}\right)^{-1}$ for $\delta=\frac{d}{\alpha}=1$ is 
\begin{equation}\label{eq:ps}
\begin{aligned}
&P_s=\text{Pr}\left(\frac{|h|^2}{1+r^{\alpha}} \geq \theta\right)\\
&=e^{-\theta} T! \left(\sum\limits_{i=0}^{T-1} \frac{(-1)^i}{(\theta R^d)^{i+1} (T-1-i)!} -
\frac{(-1)^{T-1} e^{-\theta R^d}}{(\theta R^d)^T}\right).
\end{aligned}
\end{equation}
\end{proposition}
\begin{IEEEproof} The proof has two main logical steps. First we extend the result, originally reported in \cite{finite} for distribution of distances to points of a BPP, to density function of $\alpha$-th powers of distances. Next, the latter is used to derive the probability of success of communication of a node with the nearest transmitter, taking into account both Rayleigh fading and path loss via a compound random variable.
\subsubsection{General distribution of distances and $\alpha$-th powers of distances}
We start with a general BPP, i.e. with reference point located arbitrarily (eg. Figure 1 in \cite{finite}). Under reference point we understand a receiver, and the points of the BPP are the $T_{i-1}$ transmitters. We will denote $T_{k-1}$ as $T$ for brevity.

Let $r_n$ denote the random distance from a reference point $x$ to $n$-th nearest neighbor, then, conditioned on having exactly $T$ nodes, the complementary cumulative distribution function (CCDF) of $r_n$ is \cite{finite} 
\begin{align}\label{eq:dist_BPP}
	\bar{F}_{r_n}(r)=\sum\limits_{i=0}^{n-1} {T \choose i} p^i (1-p)^{T-i},
\end{align}
where $p$ is the probability that a node falls into a subset $B$ of the observation window $W$. In case $B=b_d(x,r)$ is a $d$-dimensional ball with radius $r$ and centered at $x$, $p$ can be expressed in general in terms of counting measures as \cite[p.24]{vere}
\begin{align}
	p=p(x,r)=\frac{\Lambda(B)}{\Lambda(W)}=\int\limits_{b_d(x,r) \cap W} f(r) \text{d} r,
\end{align}
where $f(r)$ is the common density function of i.i.d. nodes randomly distributed  in $W$.
\begin{align}
	f(r)=\frac{\lambda(r)}{\Lambda(W)}=\frac{d r^{d-1}}{R^d}, r \in [0,R].
\end{align}

In case of the nearest neighbor, the CCDF and probability density function (PDF) of $r_1$ are respectively
\begin{align}\label{eq:CDF_PDF1}
	\bar{F}_{r_1}(r)&=(1-p)^{T},\\
	f_{r_1} (r)&=T(1-p)^{T-1}\frac{\text{d}p}{\text{d}r}.
\end{align}

With the assumption of an isotropic BPP and the observation point located in the origin $o$, the intersection of $b_d(o,r)$ and $W$ has an area of $b_d(o,r)$. Therefore we can express $p$ as
\begin{align}\label{eq:p}
	p=p(0,r)=\frac{\Lambda(b_d(o,r))}{\Lambda(W)}=
	\left(\frac{r}{R}\right)^d
\end{align}
Substituting this into \eqref{eq:CDF_PDF1} yields
\begin{align}
	\bar{F}_{r_1}(r)&=(1-\left(\frac{r}{R}\right)^d)^{T}=\frac{1}{R^{d T}}\left(R^d-r^d\right)^{T},\\
	f_{r_1}(r)&=T(1-p)^{T-1}\frac{\text{d}}{\text{d}r} \left(\frac{r}{R}\right)^d \nonumber \\ 
	&=\frac{T}{R^{d T}}\left(R^d-r^d\right)^{T-1}d r^{d-1}.
\end{align}
Using derived distributions property \cite[p.208]{Bert08} PDF of $r_1^{\alpha}$ can be expressed as 
\begin{equation}\label{eq:pdf_iso}
\begin{aligned}
	f_{r_1^{\alpha}}(y)&= \frac{\delta T }{R^{d T}}\left(R^d-y^\delta\right)^{T-1} y^{\delta-1}, y \in [0,R^{\alpha}].
\end{aligned}
\end{equation}

\subsubsection{The joint distribution}
We can now use the distribution of compound RV $\frac{|h|^2}{1+r^{\alpha}_1}$, since each individual distribution is known.
\begin{equation}\label{eq:Pz}
\begin{aligned}
\text{Pr}&\left(\frac{|h|^2}{1+R^{\alpha}_1} \geq \theta\right)=\int\limits_{0}^{R^{\alpha}} f_{r^{\alpha}_1}(y) \int\limits_{\theta (1+y)}^{\infty} f_{|h|^2}(x) \text{d}x \text{d}y\\
&=\frac{\delta T e^{-\theta} }{R^{d N}} \int\limits_{0}^{R^{\alpha}} \left(R^d-y^\delta\right)^{T-1} y^{\delta-1} e^{-\theta y} \text{d}y.
\end{aligned}
\end{equation}
For the special case of $\delta=1$, we get $ p=y/R^d$ and
\begin{equation}
\begin{aligned}
\text{Pr}(\frac{|h|^2}{1+r^{\alpha}_1} \geq \theta)=\frac{T e^{-\theta}}{R^{dT}} \int\limits_{0}^{R^{\alpha}=R^d} \left(R^d-y\right)^{T-1} e^{-\theta y} \text{d}y.
\end{aligned}
\end{equation}
Substituting $R^d-r=x$ we obtain
\begin{equation}
\begin{aligned}
\text{Pr}&(\frac{X}{Y} \geq \theta)=\frac{T e^{-\theta}}{R^{dT}} \int\limits_{x=R^d-0}^{x=R^d-R^d} x^{T-1} e^{-\theta (R^d-x)} (-\text{d}x)\\
&=\frac{T e^{-\theta} e^{-\theta R^d}}{R^{dT}} \int\limits_{0}^{R^d} x^{T-1} e^{\theta x} \text{d}x,
\end{aligned}
\end{equation}
Using \cite[p.176 5.1.2.1.6]{table2} we can find the solution of the integral as
\begin{equation}
\begin{aligned}
&\text{Pr}(\frac{X}{Y} \geq \theta)\\
&=\frac{T e^{-\theta}}{R^{dT}} \left(\frac{e^{-\theta (R^d-x)}}{{\theta}} \sum\limits_{i=0}^{T-1} \frac{(-1)^i}{\theta^i }\frac{(T-1)!}{(T-1-i)!} x^{T-1-i}\right)\bigg|_{x=0}^{x=R^d}\\
&=e^{-\theta} T! \left(\sum\limits_{i=0}^{T-1} \frac{(-1)^i}{(\theta R^d)^{i+1} (T-1-i)!} -
\frac{(-1)^{T-1} e^{-\theta R^d}}{(\theta R^d)^T}\right).
\end{aligned}
\end{equation}
\end{IEEEproof}
Combination of \eqref{eq:ps}, \eqref{eq:pk} and \eqref{eq:k_tx} gives us the expected number of transmissions required to reach all nodes using cooperative broadcast protocol.

\section{Numerical results and complexity analysis}
\subsection{Numerical results}
\begin{figure}[!t]
\centering
\includegraphics[width=\columnwidth]{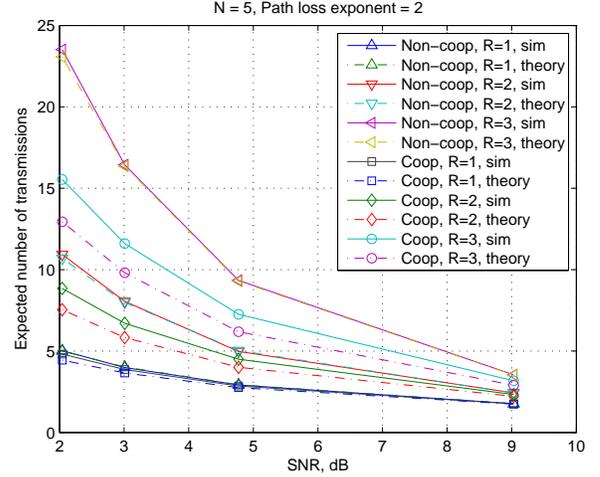}
\caption{Theoretical and simulated system latency.}
\label{fig:Number_of_retx}
\end{figure}
In this subsection we evaluate performance of cooperative and non-cooperative broadcast protocols, and verify analytical results obtained above. 
The baseline setting is as follows: $N$ nodes are placed uniformly and randomly on a disk with radius $R$, and the source of the broadcasted message is located at the center. Target data rate is 1 bit/s/Hz, channel noise is assumed to be complex Gaussian and path loss exponent $\alpha=2$. Simulation results are averaged over 1000 random network realizations.
The number of intended receivers is set to be $N=5$ and the set of radius values is $R \in [1,2,3]$.

Figure~\ref{fig:Number_of_retx} illustrates the expected number of required transmissions to complete a broadcast. We observe that simulation results for non-cooperative broadcast match tightly the analytical calculations, verifying the accuracy of the developed model. For cooperative broadcast, analytical results report lower expected system delay compared to simulations. This mismatch is understood to be caused by the assumption used in calculation of parameter $p$ in \eqref{eq:p} that the process of transmitters is isotropic with respect to any receiver. In reality, receivers closer to cell boundary have larger expected distance to the nearest transmitter. The increase of the mismatch with $R$ can be explained by the limit of integration in \eqref{eq:Pz}: for a node shifted from the origin by distance $\Delta$, the upper outer integration limit should be adjusted to $(R+\Delta)^{\alpha}$. Precise account for above factors significantly complicates analysis, whereas presented simplified model for cooperative broadcast can serve as a reasonable approximation.
\begin{figure}[!t]
\begin{center}
\subfigure[Non-cooperative broadcast]{\includegraphics[width=\columnwidth]{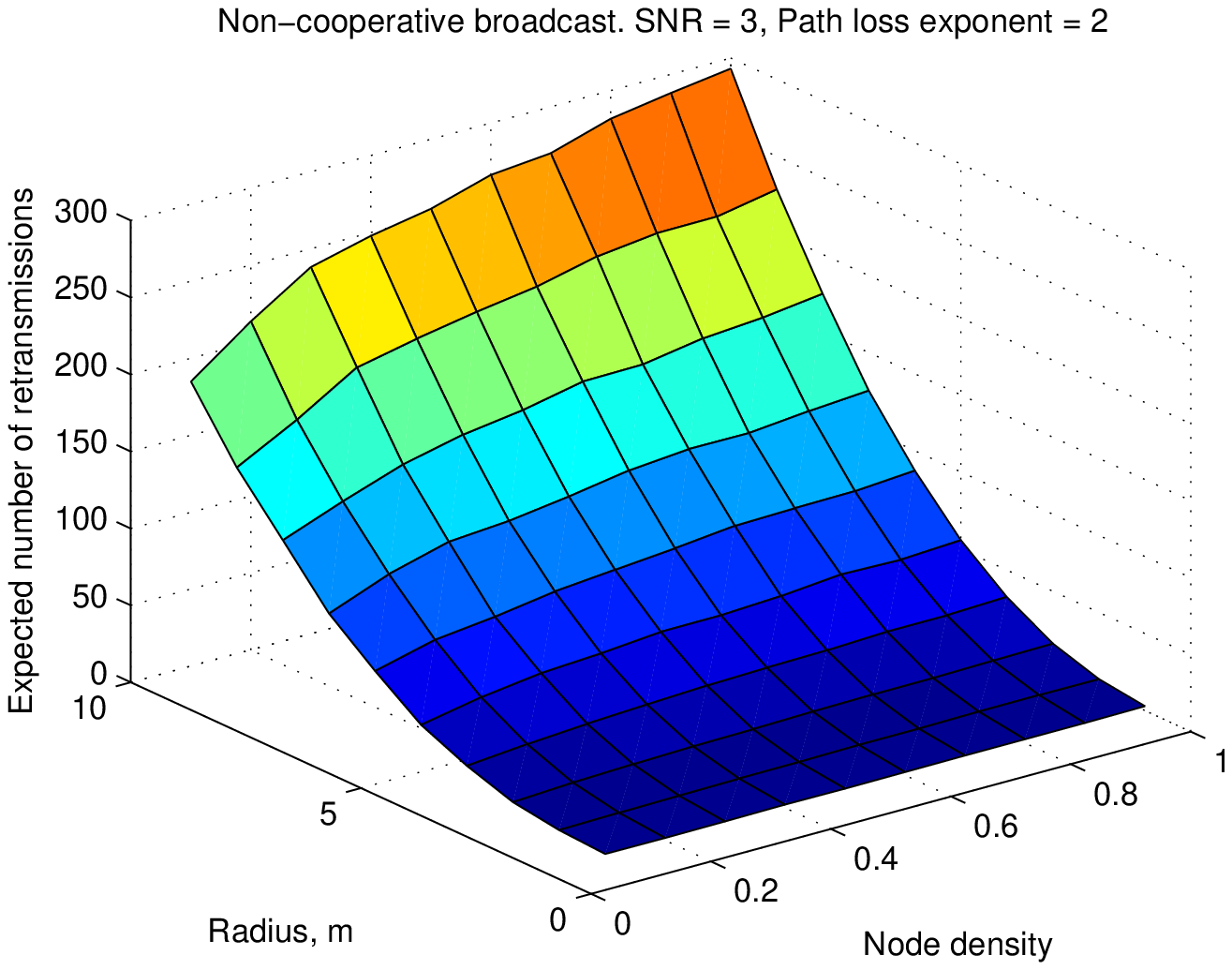}}
\subfigure[Cooperative broadcast]{\includegraphics[width=\columnwidth]{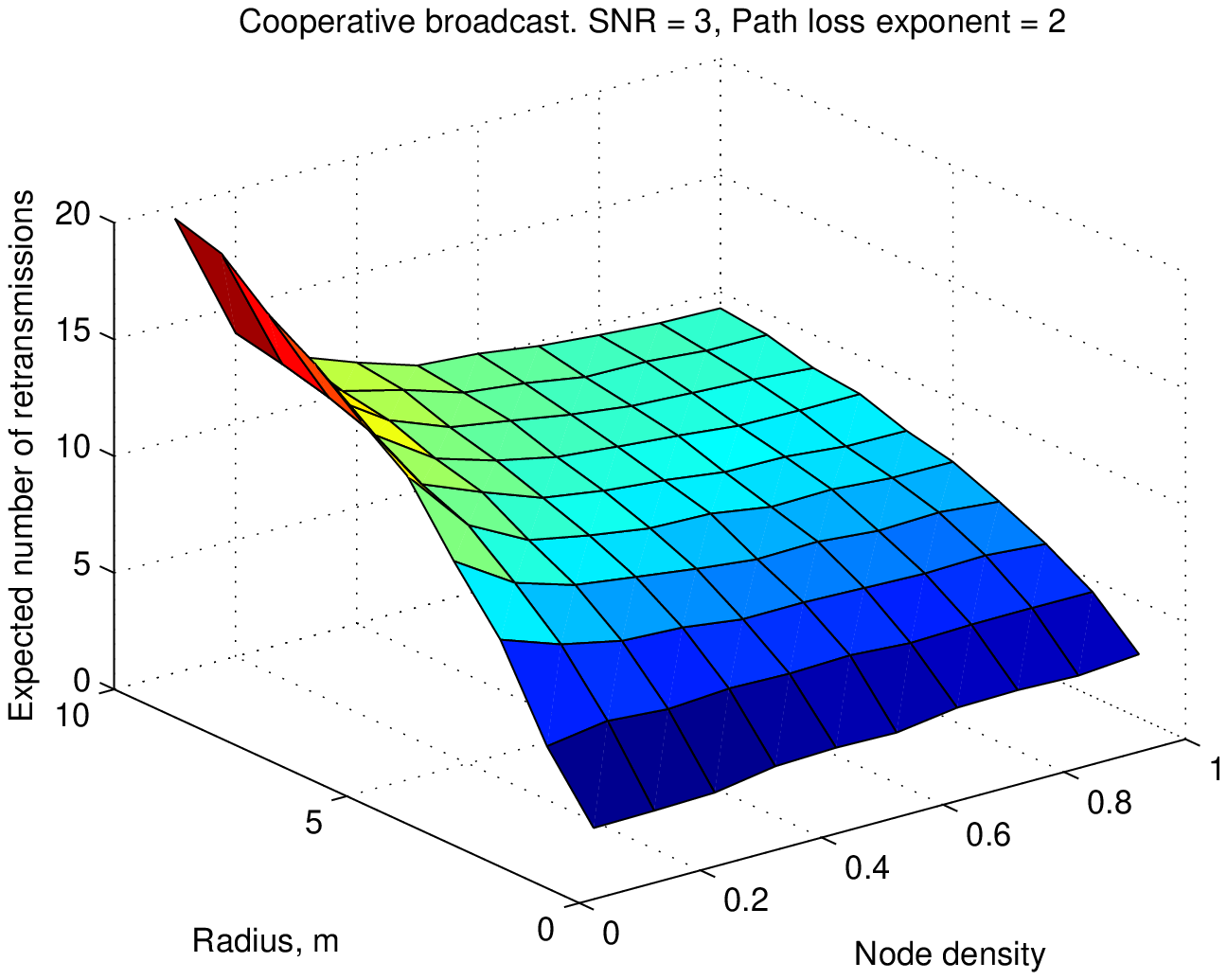}}
\caption{Transmission delay as a function of network size and node density.}
\label{fig:3d}
\end{center}
\end{figure}
From Figure~\ref{fig:Number_of_retx} we observe that cooperative broadcast achieves lower transmission delay compared to non-cooperative scenario, especially for larger-sized networks. An intuitive explanation is that in a larger network signal attenuation can be more severe for distant nodes.
To emphasize the role of network properties on studied broadcast protocols a set of simulations was performed. The SNR level was fixed at 5 dB and the number of nodes was rounded up as $N=\left\lceil \rho \pi R^2\right\rceil$, with $\rho$ denoting the node density. Figure~\ref{fig:3d} depicts the dependency of transmission delay on network size and node density for non-cooperative and cooperative schemes. It is evident that performance of the non-cooperative scheme is primarily affected by the cell size, and the node density has relatively weaker effect. 
In contrast, for cooperative scenario the impact of network size on system latency diminishes with the increase of node density.  These results suggest that cooperative schemes can be particularly effective in geometrically large networks with moderate node density. In addition, cooperative scheme achieves significantly lower delay compared to non-cooperative scheme in the same setting.

\subsection{Complexity analysis}
Presented analytical results are not in closed form and require multiple iterations of computations. In this subsection we estimate the order of computational complexity required to find the expected number of required transmissions $\bar{K}$ using \eqref{eq:numb_tx}. Note that complexities for cooperative and non-cooperative schemes differ only in calculation of the exact number of successful nodes after a transmission attempt using \eqref{eq:p1_noncoop} or \eqref{eq:pk}. Therefore, we develop the complexity order based on the non-cooperative case.

First, to obtain $\bar{K}$, summation of infinite number of terms is required in \eqref{eq:numb_tx}. However, the number of terms, contributing significantly to $\bar{K}$ can be limited by some threshold $K'$. For example, the typical number of transmissions $\bar{K}$ required to reach 10 nodes at 3 dB transmit SNR in Rayleigh fading is roughly 16, which is far below infinity and does not change significantly after $K'=25$. Therefore, we will use the following approximation:
\begin{equation}
	\bar{K}=\sum\limits_{k=1}^{\infty} k \cdot P(K=k|N) \approx \sum\limits_{k=1}^{K'} k \cdot P(K=k|N)
\end{equation}
%Note, that the summation itself does not contribute to the order of complexity, which will be determined by the largest-order term. 
We will further estimate the number of operations $X_{1}$ required to evaluate (1) as a function of the number of nodes $N$ and the threshold $K'$.

Expression \eqref{eq:k_tx_ref} is a summation of all possible outcomes of $k\in [1,K']$ transmissions leading to delivery of broadcasted message to all $N$ nodes. Using methodology in \cite[p.43]{knuth1989}, one can find that the number of such combinations is $C(N,k)=\frac{(N+k-2)!}{(N-1)!(k-1)!}$.
Each summation term in (5) consists of a product of $k$ probabilities, calculated using \eqref{eq:p1_noncoop}.
Let us denote as $X$ the number of operations (i.e. additions or multiplications), required to evaluate \eqref{eq:p1_noncoop}. Then, each term of summation in \eqref{eq:k_tx_ref} would need $X^{k}$ operations, and evaluation of \eqref{eq:k_tx_ref} would take $X_{5}=X^{k} \cdot C(N,k)$ such operations. Finally, the number of operations required to evaluate \eqref{eq:numb_tx} can be found as 
\begin{equation}\label{eq:X1}
	X_{1}=\sum\limits_{k=1}^{K'} X^{k} \cdot C(N,k) + \epsilon,
\end{equation}
where $\epsilon$ is the total number of intermediate multiplications and additions involved in \eqref{eq:numb_tx} and \eqref{eq:k_tx_ref}.
Since order of complexity is determined by the highest-order term \cite{pap82} we can ignore $\epsilon$ and express it as
\begin{equation}
	\mathcal{O} \left(X^{K'}C(N,K')\right).
\end{equation}
Such complexity order restricts application of this type of analysis to large values of system parameters, however for finite networks with fixed and small number of nodes, calculations can be done in realistic times.

\section{Conclusion}
In this correspondence, we analyzed the impact of node cooperation on system delay for broadcast scenarios in finite wireless networks. Developed analytical models, that incorporate both channel and node location randomness, match precisely simulated non-cooperative broadcast, and provide a close approximation for cooperative scheme. The framework used in this work can be extended to capturing important network geometry properties in more complex scenarios.
\IEEEtriggeratref{4}

\end{document}